\documentclass[11pt]{article}
\usepackage{moriond,epsfig}

\newcommand{\apj}{{\it ApJ}}

\newcommand{\apjl}{{\it ApJ}}
\newcommand{\aap}{{\it A\&A}}

\newcommand{\aapr}{{\it A\&A~Rev}}
\newcommand{\mnras}{{\it MNRAS}}

\newcommand{\nat}{{\it Nature}}

\def\be{\begin{equation}}
\def\ee{\end{equation}}
\def\bea{\begin{eqnarray}}
\def\eea{\end{eqnarray}}

\def\sedici {SGR\,1627--41}
\def\sgr {1E\,1547.0--5408}
\def\pdot {\dot P}

\begin{document}
\vspace*{4cm}
\title{RECENT RESULTS ON MAGNETARS}

\author{ SANDRO MEREGHETTI }

\address{INAF - Istituto di Astrofisica Spaziale e Fisica Cosmica Milano, \\
v. E.Bassini
15, I-20133 Milano, Italy}

\maketitle\abstracts{Several observations obtained in the last few
years indicate that Soft Gamma-ray Repeaters (SGRs) and Anomalous
X-ray Pulsars (AXPs) are basically a single class of isolated
neutron stars. Their properties are well explained by the magnetar
model, based on neutron stars powered by magnetic fields as high
as 10$^{14}$--10$^{15}$ G.  Here I report some recent results
obtained for the transient Soft Gamma-ray Repeater  \sedici , that
started a new outburst after about 10 years from the previous one,
and for  the Anomalous X--ray Pulsar \sgr . The latter source
recently showed a remarkable bursting activity, that reinforces
the similarity between AXPs and SGRs.}

\section{Anomalous X-ray Pulsars and Soft Gamma-Ray Repeaters}

Two small classes of peculiar high-energy sources, the Anomalous
X-ray Pulsars (AXPs) and the Soft Gamma-Ray Repeaters (SGRs), have
attracted increasing attention in the last decade. Their
classification in two distinct groups reflects the different
manifestations that led to their discovery, but there is mounting
evidence that they are probably a single class of objects. Namely,
observations performed in the last few years showed many
similarities between AXPs and SGRs \cite{woo06,mer08}.

AXPs were first detected as persistently bright pulsars in the
soft X-ray range ($<$10 keV) and thought to belong to the
population of galactic accreting binaries. When more X--ray data
accumulated, and deeper optical/IR searches excluded the presence
of bright companion stars, their peculiar properties started to
emerge and led to classify them  as a separate class of pulsars
\cite{mer95}. Their common properties are periods of a few
seconds, secular spin-down in the range 10$^{-13}$ -- 10$^{-10}$ s
s$^{-1}$, relatively soft spectra below 10 keV, and, in some
cases, associations with supernova remnants.

SGRs were discovered through the detection of short bursts in the
hard X-ray/soft gamma-ray range and initially considered as a
subclass of gamma-ray bursts  \cite{lar86,att87}. During sporadic
periods of activity, lasting from days to months, they emit short
bursts ($<$1 s) of hard X--rays/soft $\gamma$-rays reaching
L$\sim$10$^{41}$ erg s$^{-1}$. Occasionally, SGRs emit much more
energetic ``giant flares'' with luminosity  up to 10$^{47}$ erg
s$^{-1}$. Only three of these rare events have been observed, each
one from a different source~\cite{maz79,hur99,pal05,mer05b}. When
good positions for the SGRs bursts could be obtained it became
possible to identify their X-ray counterparts, finding that they
are pulsating sources very similar to the AXPs.

It is generally believed that both SGRs and AXPs are Magnetars:
neutron stars with extremely high magnetic fields,
B$\sim$10$^{14}$-10$^{15}$ G,  i.e.  100-1000 times stronger than
those of the typical neutron stars observed as radio pulsars
powered by  rotational energy or   as X--ray pulsars powered by
accretion from their binary companions. In this interpretation,
the magnetic field is the ultimate energy source of all the
persistent and bursting emission observed  in AXPs and SGRs
\cite{tho95,tho96}.

\begin{figure}
%\rule{5cm}{0.2mm}\hfill\rule{5cm}{0.2mm} \vskip 2.5cm
%\rule{5cm}{0.2mm}\hfill\rule{5cm}{0.2mm}
\begin{center}
\psfig{figure=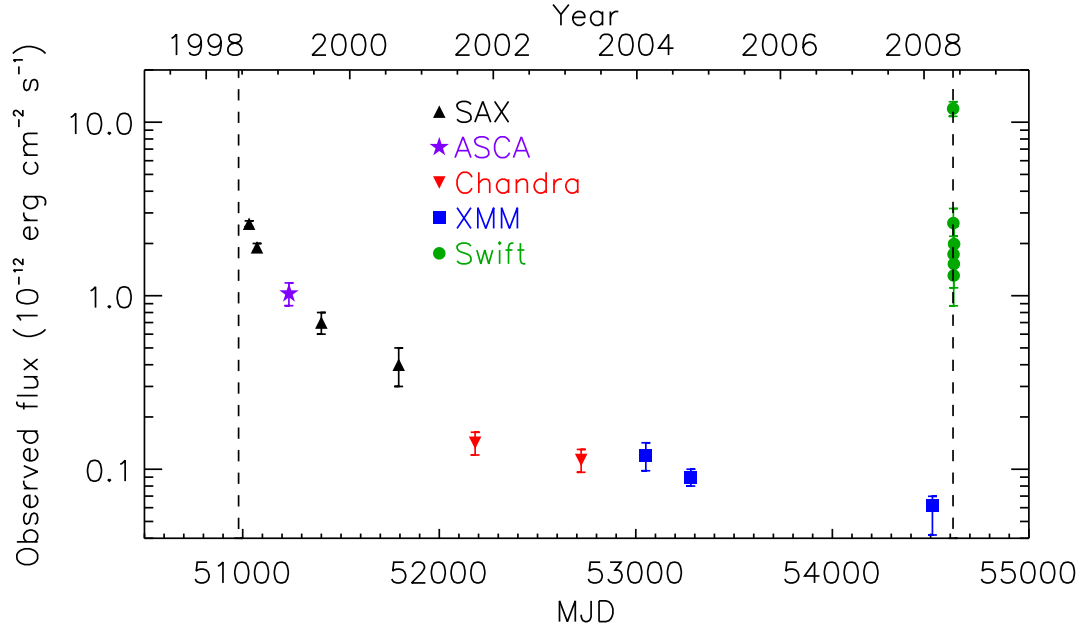,height=9cm} \caption{X--ray light curve of
\sedici\   spanning ten years of observations with different
satellites (observed flux in the 2-10 keV energy range).  The
vertical lines indicate the two periods of bursting activity seen
from this source (June 1998 and May 2008). \label{fig:lc1627}}
\end{center}
\end{figure}

\begin{figure}
%\rule{5cm}{0.2mm}\hfill\rule{5cm}{0.2mm} \vskip 2.5cm
%\rule{5cm}{0.2mm}\hfill\rule{5cm}{0.2mm}
\begin{center}
 \psfig{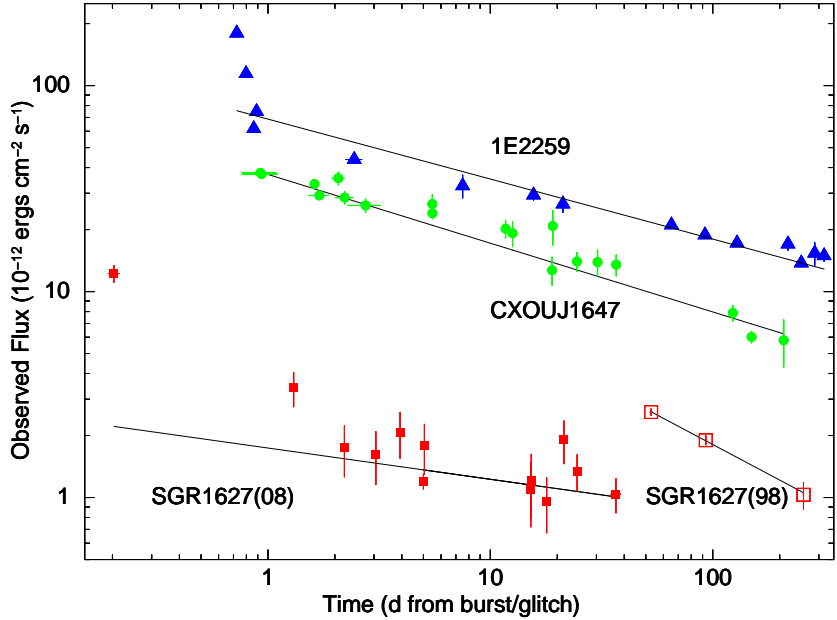}
 \caption{Comparison of the long term flux decays following outbursts
of SGRs and AXPs. % (from \cite{esp08}).
For \sedici\ both the 1998 and 2008 events are plotted.  Note the
good coverage of the early phases of the 2008 outburst that could
be obtained thanks to the prompt observations with Swift. The
lines are power laws with time decay index ranging from --0.2
(\sedici\ in 2008) to --0.6 (\sedici\ in 1998).
\label{fig:decays}}
\end{center}
\end{figure}

Here I   present some results on two sources that entered new
periods of strong activity in the last months: the \sedici\ and
the AXP \sgr . These results further support the similarity
between these two classes of sources.

\section{The 2008 reactivation of SGR 1627--41}

\sedici\ was the first  SGR to show a  transient behavior. It was
discovered in 1998, during a bursting state that lasted about six
weeks~\cite{woo99}. At the time of the outburst its X-ray
counterpart had a luminosity of $\sim 10^{35}$ erg s$^{-1}$, but
in the following years its X--ray luminosity gradually
decreased~\cite{mer06a}, as shown in Fig.~\ref{fig:lc1627}. This
long-term decay was interpreted as the cooling of the neutron star
after the heating that occurred during the outburst~\cite{kou03}.
In principle, the modelling of the long term light curve could
provide information on the mechanism for (and location of) the
heating and on the neutron star structure. However, uncertainties
in the relative cross calibrations of the different instruments as
well as the limited spectral information make it difficult to
obtain reliable results~\cite{mer06a}.

An XMM-Newton observation carried out in February 2008 revealed
\sedici\ at only $\sim10^{33}$ erg s$^{-1}$ (for d=11 kpc), the
lowest luminosity observed for a SGR~\cite{esp08}. In May 2008
\sedici\ started a new outburst, during which several short bursts
were detected and a peak luminosity higher than that observed in
1998 was attained (Fig.~\ref{fig:lc1627}). The subsequent
evolution could be monitored by a series of Swift observations
that showed an initial rapid decrease followed by a shallower
phase~\cite{esp08}. The 2008 light curve is compared in
Fig.~\ref{fig:decays} with that of the previous outburst and with
the behavior seen in a few other AXPs/SGRs~\cite{esp08}. When
early data are available, they show that a single power law decay
cannot reproduce the source fading, owing to the presence of a
steeper initial phase in the first days after the outburst. This
suggest the presence of two different mechanisms at play. One
possibility is that the steep phase be due to magnetospheric
currents dissipation while the later phase reflect the effect of
crustal cooling. It is also possible that X-rays emitted during
the initial bright burst, delayed by interstellar dust scattering,
contribute to the initial steep phase (see below).

\begin{figure}
\begin{center}
%\rule{5cm}{0.2mm}\hfill\rule{5cm}{0.2mm} \vskip 2.5cm
%\rule{5cm}{0.2mm}\hfill\rule{5cm}{0.2mm}
\psfig{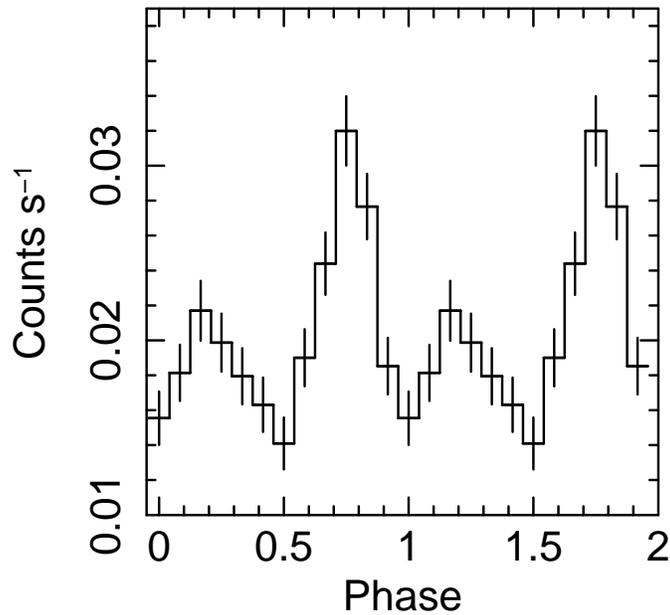}
 \caption{X-ray light curve of \sedici\ folded at the spin period of 2.59
s discovered by Esposito et al. (2009). The data have been
obtained with the XMM-Newton EPIC pn camera in the 2-12 keV energy
range. \label{fig:fol1627}}
\end{center}
\end{figure}

Due to visibility constraints, the brightest part of the \sedici\
outburst could not be observed by XMM-Newton, but we requested a
Target of Opportunity observation to be performed as soon as
possible, in order to take advantage of the source brightness to
measure its still unknown spin period. The observation was done on
2008 September 27-28, and despite the low source flux
$\sim3\times10^{-13}$ erg cm$^{-2}$ s$^{-1}$, the large effective
area of the EPIC instrument allowed us to collect enough counts
and to discover the long-sought pulsations \cite{esp09}. The spin
period is  2.6 s, one of the shortest among magnetar candidates.
The X--ray pulse profile, characterized by two peaks of different
intensity, is shown in Fig.~\ref{fig:fol1627}.

\begin{figure}
\begin{center}
%\rule{5cm}{0.2mm}\hfill\rule{5cm}{0.2mm} \vskip 2.5cm
%\rule{5cm}{0.2mm}\hfill\rule{5cm}{0.2mm}
\psfig{figure=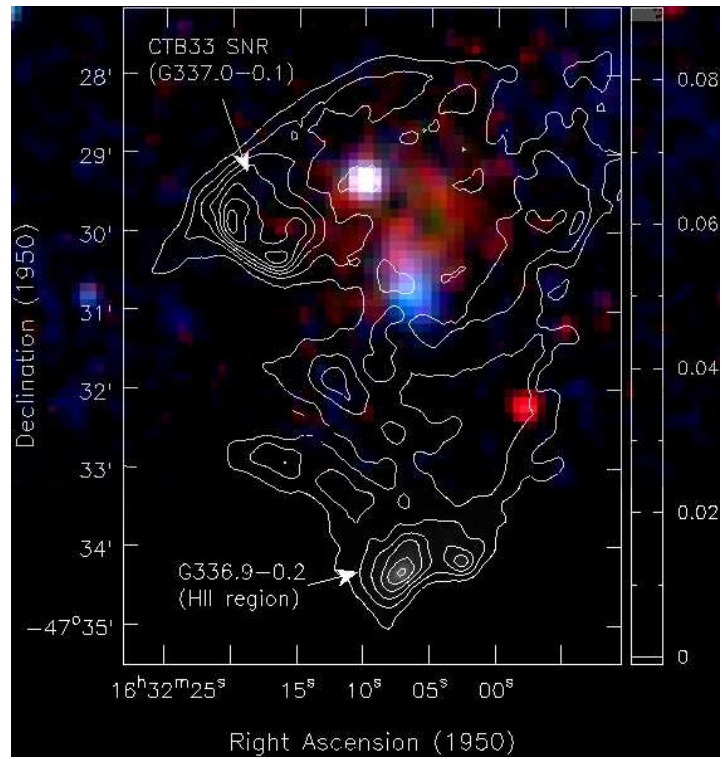,height=10cm}
 \caption{XMM-Newton EPIC X-ray image of the region of \sedici\
with overlaid contours from the 1375 MHz radio map of Sarma et al.
(1997). The colors indicate the photon energy (1.7--3.1 keV in
red, 3.1--5 keV in green, and 5--8 keV in blue). The bright source
in white is \sedici . The bluish diffuse source is most likely a
cluster of galaxy in the background, as indicated by its high
absorption and redshifted Fe line. The soft X--ray (in red)
diffuse emission can be associated to the SNR G337.0--0.1.
\label{fig:ima1627}}
\end{center}
\end{figure}

The deep XMM-Newton observation led also to the discovery of
diffuse X--ray emission from the vicinity of \sedici , as shown in
Fig.~\ref{fig:ima1627}.  Spectral and spatial analysis shows that
the resolved source about 1.5 arcmin south of the SGR is most
likely due to a cluster of galaxies, while the more extended and
softer emission is related to the  supernova remnant / HII region
complex CTB~33~\cite{sar97}.

\vspace{0.5cm}

\section{The January 2009 outburst of 1E 1547.0--5408}

\begin{figure}
\begin{center}
%\rule{5cm}{0.2mm}\hfill\rule{5cm}{0.2mm} \vskip 2.5cm
%\rule{5cm}{0.2mm}\hfill\rule{5cm}{0.2mm}
\psfig{figure=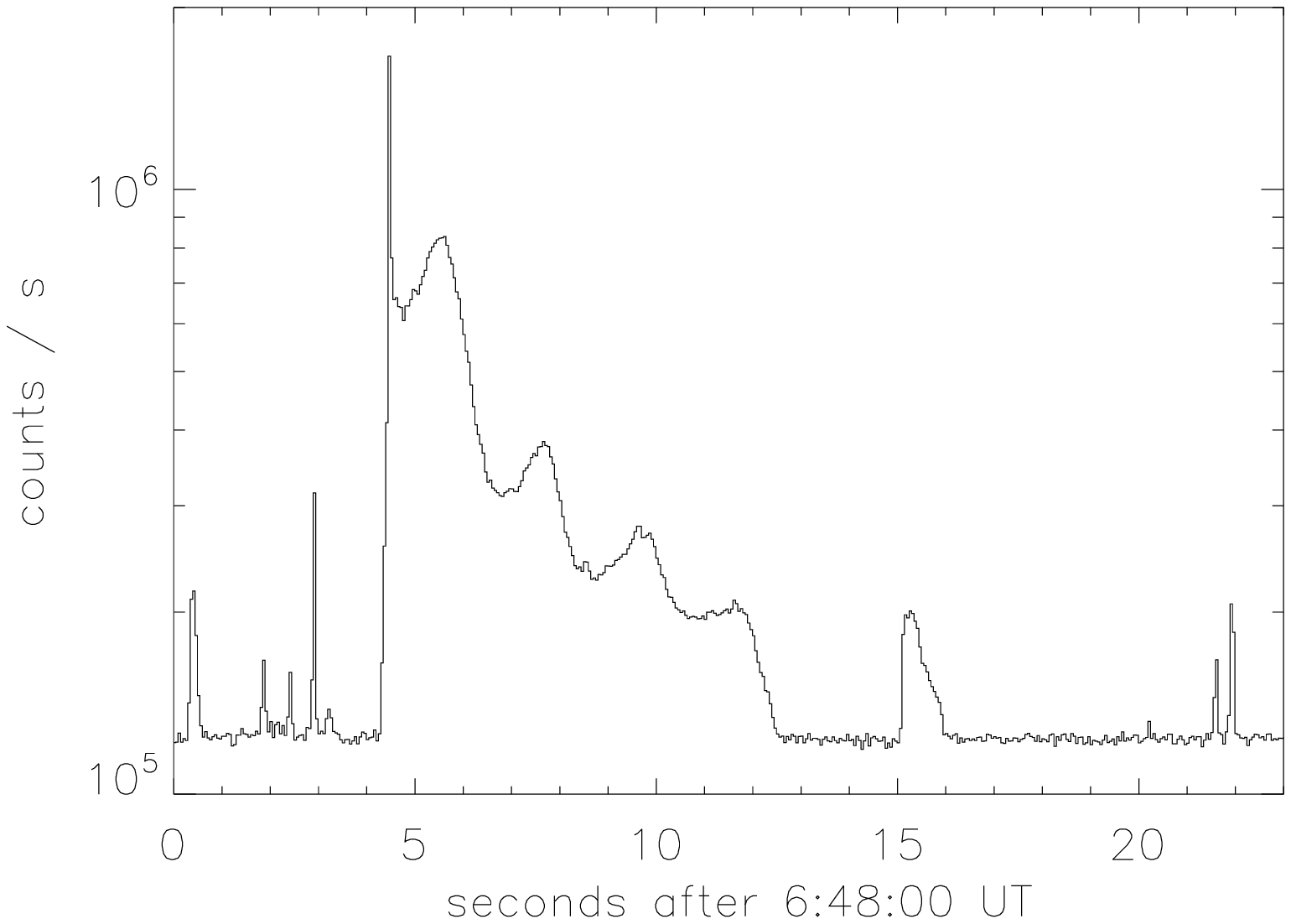,height=9cm}
 \caption{Bursts from \sgr\ observed at E$>$80 keV with the
Anti-Coincidence System of the SPI instrument on board INTEGRAL on
January 22, 2009. The initial spike of the longest burst had a
duration of $\sim$0.3 s and reached a peak flux greater than 2
10$^{-4}$ erg cm$^{-2}$ s$^{-1}$ (25 keV - 2 MeV). A modulation at
2.1 s,  reflecting  the neutron star rotation period,  is clearly
visible in the burst tail.
  \label{fig:burst1547}}
\end{center}
\end{figure}

\begin{figure}
\begin{center}
%\rule{5cm}{0.2mm}\hfill\rule{5cm}{0.2mm} \vskip 2.5cm
%\rule{5cm}{0.2mm}\hfill\rule{5cm}{0.2mm}
\psfig{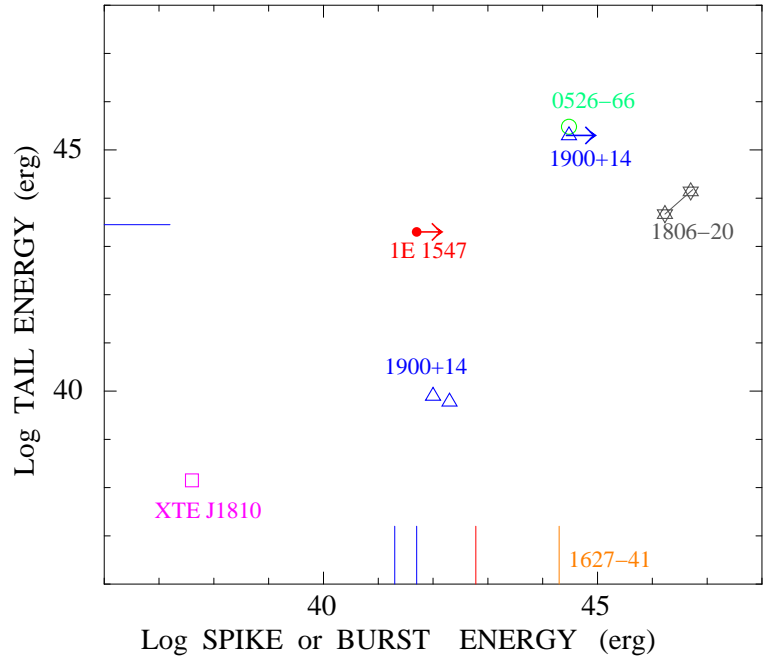}
 \caption{Energetics of flares and peculiar bursts from SGRs and
AXPs. The different sources are distinguished by the symbols
color. The ordinate gives the energy in the pulsating tails that
often follow the brightest bursts, while the abscissa reports the
energy in the initial spikes (data from Mereghetti et al. (2009)
and references therein). The vertical/horizontal lines refer to
events in which only one of these components has been observed.
The three historical giant flares from SGRs are in the upper right
corner. Note that in some cases only lower limits to the total
energy could be derived due to instrument saturation. The two
points for SGR 1806--20 are for the generally assumed distance of
15 kpc and for the more recent estimate d=8.7 kpc. The energetics
of the burst from \sgr\, for an assumed distance of 10 kpc, is in
the range of the so called ``intermediate flares''.
  \label{fig:ene}}
\end{center}
\end{figure}

The transient X--ray source \sgr\ was discovered almost 30 years
ago~\cite{lam81} in the supernova remnant G\,327.24--0.13, but it
attracted little interest until it was proposed as a possible AXP
on the basis of new X-ray and optical studies ruling out more
standard interpretations~\cite{gel07}. Radio pulsations with
$P=2.1$ s and period derivative $\pdot=2.3\times10^{-11}$ s
s$^{-1}$ were subsequently discovered~\cite{cam07c}, confirming
its AXP nature. In October 2008 \sgr\ started an outburst with the
emission of several short bursts and a significant increase in its
X-ray flux~\cite{isr09}.

More recently, a new period of strong activity culminated on 2009
January 22, when more than 200 bursts were detected in a few
hours. Some of these bursts were particularly bright, and two had
durations sufficiently long to show a clear modulation at the
neutron star spin period. Of particular interest is the burst
shown in Fig.~\ref{fig:burst1547}, that started with a very bright
and short initial spike ($\sim$0.3 s) followed by a $\sim$8 s long
pulsating tail~\cite{mer09}. Although these features  are typical
of giant flares from SGRs \cite{maz79,hur99,pal05,mer05b}, the
energy released in this event was not as large as that of the
three historical giant flares. This is shown in
Fig.~\ref{fig:ene}, where the energetics of the strongest bursts
and flares from SGRs/AXPs are compared. Although the plotted data
are affected by some uncertainties, especially for the brightest
events that caused instrument saturation, it is clear that there
is a rather continuum distributions of intensities, from the
normal short bursts up to the brightest giant flares.  It is also
noteworthy that extended pulsating tails have been detected not
only for the three giant flares, but also after less intense
bursts~\cite{ibr01,len03,woo05}. Conversely, also a few examples
of pulsating tails apparently without a bright initial hard spike
have been observed~\cite{gui04,mer09}. This is possibly an
indication that the spike emission is non-isotropic, a fact that
adds a further uncertainty to proper estimates of the involved
energy.

\begin{figure}
\begin{center}
%\rule{5cm}{0.2mm}\hfill\rule{5cm}{0.2mm} \vskip 2.5cm
%\rule{5cm}{0.2mm}\hfill\rule{5cm}{0.2mm}
\psfig{figure=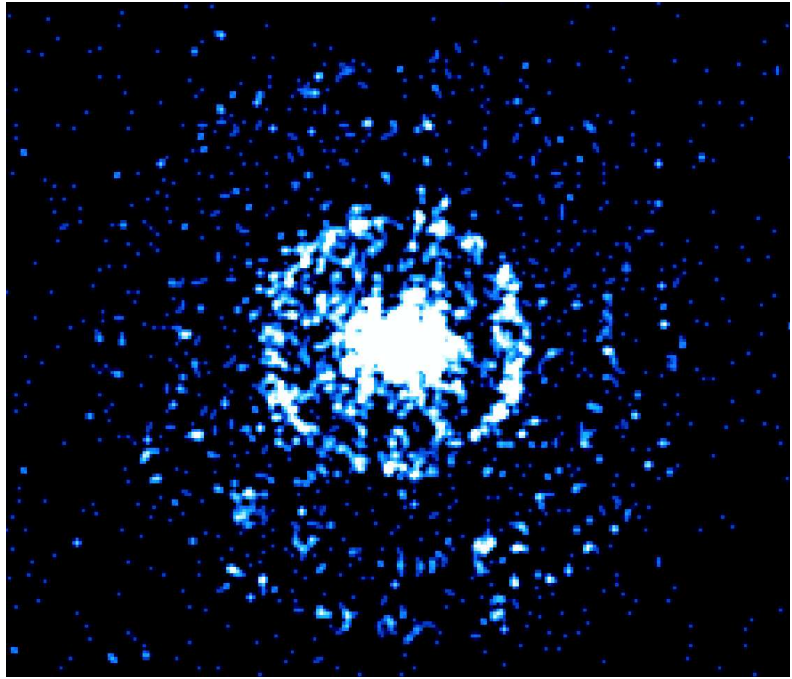,height=9cm}
 \caption{X-ray rings produced by dust scattering around \sgr .
In this image, obtained with the Swift/XRT instrument on January
23, the innermost and brightest ring has a radius of $\sim$1
arcmin. Two outer rings, produced by closer dust layers are also
visible. The ring dimensions were seen to increase in later
observations, as expected for scattering by narrow dust layers of
the X--ray flux emitted during the strong bursting activity that
took place around 6:48 UT of January 22. \label{fig:ring}}
\end{center}
\end{figure}

Immediately after the discovery of the strong bursting activity of
January 22, several follow-up pointings of \sgr\ were carried out
with Swift. During the first XRT observations, the imaging mode
could not be used because the source was too bright. The first
data providing full imaging (Fig.~\ref{fig:ring}) were obtained on
January 23 at $\sim$15:30 UT and showed the presence of remarkable
dust scattering rings around the source position \cite{tie09}.
Dust scattering X-ray halos around bright galactic sources were
predicted well before their observations with the first X-ray
imaging instruments~\cite{ove65}. Their study allows to get
information on the properties and spatial distribution of the
interstellar dust. When the scattered radiation is a short
burst/flare and the dust is concentrated in a relatively narrow
cloud, an expanding ring (instead than a steady diffuse halo)
appears, due to the difference in path-lengths at different
scattering angles. Halos in the form of expanding rings have been
observed in a few gamma-ray bursts, and their study allowed to
determine accurate distances for the scattering dust clouds  in
our galaxy~\cite{vau04,tie06,via07}.

The dust scattering rings around \sgr\ are the brightest ever
observed and the first ones for an AXP/SGR. Further observations
carried out with Swift, XMM-Newton and Chandra clearly show that
their angular size is increasing with time. By fitting their
expansion law it is possible to determine the burst emission time,
which is found to coincide with the interval of highest activity
including the bright event of $\sim$6:48 UT shown in
Fig.~\ref{fig:burst1547}. A comprehensive spectral analysis of all
the available X--ray data of the expanding rings around \sgr\ will
allow to determine the distances of the source and of the three
dust layers~\cite{tie09b}.

\section{Conclusions}

The two sources described here have many similarities. Their spin
periods (2.1 and 2.6 s) are the shortest of all the AXPs/SGRs,
both are  located in supernova remnants (and are in the same
region of the galactic plane), both are transient sources and
emitted short bursts when in the high state. If \sgr\ had not been
previously known as a weak X-ray source, but discovered through
its bursts it would have been baptized SGR, as it recently
happened for the new source  SGR 0501$+$4516~\cite{eno09,rea09}.
This underlines once more that the distinction between AXPs and
SGRs does not reflect a real difference in the two classes of
sources, that can be well explained by the same physical model.

\section*{Acknowledgments}
The results described in this work have been possible thanks to a
large international collaboration including  P.Esposito, A. De
Luca,  D.G\"{o}tz, G.L.Israel,  N.Rea, L.Stella,  A.Tiengo,
R.Turolla, G.Vianello, A. von Kienlin, G.Weidenspointner, S. Zane
and others.

\end{document}